\documentclass[%
 reprint,
superscriptaddress,
 amsmath,amssymb,
 aps,
]{revtex4-2}

\usepackage{aps_macros}
\usepackage{pifont}
\usepackage{graphicx}
\usepackage{dcolumn}
\usepackage{bm}
\usepackage{hyperref}

\usepackage{subfigure}%
\usepackage{float}%
\hypersetup{
     colorlinks   = true,
     citecolor    = blue
}

\hypersetup{colorlinks=true, citecolor=blue, linkcolor = magenta}

\newcommand{\xmark}{\ding{55}}

\begin{document}

\preprint{APS/123-QED}

\title{Realistic Anisotropic Neutron Stars: Pressure Effects}

\author{L.~M.~Becerra}
\email{laura.becerra7@correo.uis.edu.co}
\affiliation{Grupo de Investigaci\'on en Relatividad y Gravitaci\'on, Escuela de F\'isica, Universidad Industrial de Santander A. A. 678, Bucaramanga 680002, Colombia}
 
\author{E.~A.~Becerra-Vergara}
\email{eduar.becerra@correo.uis.edu.co}
\affiliation{Grupo de Investigaci\'on en Relatividad y Gravitaci\'on, Escuela de F\'isica, Universidad Industrial de Santander A. A. 678, Bucaramanga 680002, Colombia}

\author{F.~D.~Lora-Clavijo} 
\email{fadulora@uis.edu.co}
\affiliation{Grupo de Investigaci\'on en Relatividad y Gravitaci\'on, Escuela de F\'isica, Universidad Industrial de Santander A. A. 678, Bucaramanga 680002, Colombia}

\date{\today}

\begin{abstract}
 In this paper, we study the impact of anisotropy on neutron stars with different equations of state,  which have been modeled by a piecewise polytropic function with continuous sound speed. Anisotropic pressure in neutron stars is often attributed to interior magnetic fields, rotation, and the presence of exotic matter or condensates. We quantify the presence of anisotropy within the star by assuming a quasi-local relationship. We find that the radial and tangential sound velocities constrain the range of anisotropy allowed within the star. As expected, the anisotropy affects the macroscopic properties of stars, and it can be introduced to reconcile them with astrophysical observations. For instance, the maximum mass of anisotropic neutron stars can be increased by up to 15\% compared to the maximum mass of the corresponding isotropic configuration. This allows neutron stars to reach masses greater than $2.5M_\odot$, which may explain the secondary compact object of the GW190814 event. Additionally, we propose a universal relation for the binding energy of an anisotropic neutron star as a function of the star's compactness and the degree of anisotropy.
\end{abstract}

\maketitle


\section{\label{sec:intro} Introduction}

Neutron stars (NS) are extremely dense and compact objects that have traditionally been modeled as isotropic configurations \citep{2000ARNPS..50..481H,2012ARNPS..62..485L,2016ARA&A..54..401O,2016ApJ...820...28O}. Recently, researchers have focused on  using  anisotropic fluids to provide a more accurate description 
of their internal structure, composition, and to account for  various physical phenomena associated with them 
~\citep{1997PhR...286...53H,1974ApJ...188..657B,1981JMP....22..118C,1975A&A....38...51H,PhysRevD.80.064039,PhysRevD.77.027502,2015Ap&SS.359...13B,2019PhRvD.100j3006B,2021MPLA...3650028B,2023EPJC...83..307B}. For example, phenomena such as phase transitions \citep{sokolov1980phase}, pion condensation \citep{1972PhRvL..29..382S},  the presence of very high magnetic field \citep{2012MNRAS.427.3406F,2015MNRAS.447.3278B}, and relativistic nuclear interactions \citep{1975ARA&A..13..335C} could produce anisotropy in the stellar interior, mainly where nuclear matter reaches densities greater than $10^{15}$ g/cm$^3$ \citep[see][for a complete review of these mechanisms and of the anisotropic compact stellar solution in general relativity]{1972ARA&A..10..427R, 1997PhR...286...53H,2003GReGr..35.1435D,2023EPJC...83..307B}.

The study of anisotropic pressure effects on NS can shed light on the behavior of matter under extreme conditions and provide insights into the equation of state (EOS), which describes the relationship between the pressure, density, and composition of matter. The anisotropic distribution of nuclear matter can reveal 
 the presence of exotic particles~\citep{PhysRevLett.29.382} such as hyperons, kaon condensates, or a deconfined phase of strange matter \citep{1997PhRvC..55.1587B}, as well as phase transitions~\citep{1998NuPhB.531..478C}, the crystallization of the core \citep{2012PhRvD..85l3004N}, superfluid core \citep{2000PhR...328..237H}, and the nature of the strong force interactions~\citep{2015PhLB..742..136A,2015PhRvC..92b5802A,2019PhRvD.100j3006B}. This knowledge is crucial for understanding the behavior of matter in the early universe, nuclear physics, and astrophysical processes involving high densities.

The first work exploring the impact of anisotropic pressure on stable NS configurations was carried out by \citet{1974ApJ...188..657B}. They requested that the anisotropy vanishes at the origin, and assumed that it exhibits a non-linear dependence on radial pressure, and is induced by gravity. Their findings indicated that this anisotropy has non-negligible effects on the mass and redshift of an NS. Since then, numerous publications have extensively studied solutions of Einstein’s equations for spherically symmetric static arrangements, considering anisotropic pressure \citep{1975A&A....38...51H,1981JMP....22..118C,2011IJMPD..20..319I,2011CQGra..28b5009H,2002JMP....43.4889H,2003GReGr..35.1435D,2000astro.ph.12265D}. These works reveal that anisotropy has remarkable effects on the structure and properties of NS, including observable properties such as the mass-radius ratio \citep{1974ApJ...188..657B,2021ApJ...922..149D}, moment of inertia \citep{2020EPJC...80..769R}, redshift \citep{2021arXiv211212518K}, tidal deformability \cite{2021EPJC...81..698P,2019PhRvC.100e5804R,2019PhRvD..99j4002B}, maximum mass \citep{2000astro.ph.12265D,2021Ap&SS.366....9R}, and non-radial oscillation \citep{2012PhRvD..85l4023D}.  Furthermore, anisotropy can stabilize stellar configurations that would otherwise be unstable~\citep{2011CQGra..28b5009H,2003GReGr..35.1435D,2018EPJC...78..673E}. For example, \citet{2019PhRvD..99j4002B} noted that certain EOS ruled out by gravitational waves and electromagnetic observations could become viable if the star attains a substantial degree of anisotropy.

The influence of pressure anisotropy on stellar properties varies depending on the selected model and the amount of anisotropy present. However, these factors can be constrained by the analysis of observational data. \citet{2015CQGra..32n5008S} suggested that binary pulsar observations could constrain the range of anisotropy. \citet{2021PhRvC.104f5805R} used a relativistic mean-field model to show that an anisotropic NS, based on the \citet{2011CQGra..28b5009H} model, is consistent with the constraints imposed by NS multimessenger observations. Furthermore, 
the range of pressure anisotropy inside stars was constrained in \citep{2019PhRvD..99j4002B} using tidal deformability data from GW170817. Similarly, \citet{2022PhRvD.106j3518D} and \citet{2023arXiv230515724R} constrained the canonical moment of inertia and f-mode frequency for different degrees of anisotropy using the GW170817 and GW190814 events. Additionally, the inferred mass and equatorial radius from the NICER observations of the pulsars PSR J0030+045 \cite{2019ApJ...887L..21R,2019ApJ...887L..24M} and PSR J0740+662 \cite{2021ApJ...918L..27R,2021ApJ...918L..28M} could further constrain the anisotropy range allowed within a NS.

The significance of accounting for anisotropy within NSs is undeniable. This phenomenon holds the potential to guide us towards a more profound comprehension of the intricate interplay between the internal structure and observable properties inherent to these astrophysical entities. Therefore, our paper focuses on investigating the impact of anisotropy on the macroscopic properties of NSs, such as mass, radius, compactness, and binding energy. We obtain anisotropic NS configurations by solving Einstein's field equations for spherically symmetric matter. In this context, we characterize the anisotropy of pressure within the star by adopting the quasi-local relation proposed by \citet{2011CQGra..28b5009H}. According to this relation, the difference between the radial and tangential pressure is assumed to be proportional to the star's local compactness and radial pressure,   ensuring that anisotropy vanishes at the center of the star. Furthermore, we study different nuclear  EOSs covering different models and particle compositions of the star. We parameterize them using the Generalized Piecewise Polytropic (GPP)  fit proposed by \cite{Boyle2020}, which   guarantees continuity in radial pressure and sound speed within the star.

The paper is organized as follows. In Sec.~\ref{sec:NEM}, we present  the equations of structure for anisotropic spherically symmetric stars in General Relativity (Sec.~\ref{subsec:TOV} ). We also introduce the parameterization used to model various EOSs for the NS matter with the GPP fit (Sec.~\ref{subsec:EOS})
, and give a brief description of the numerical code used to integrate the equations (Sec.~\ref{subsec:NC}). The macroscopic properties of anisotropic NSs, including mass-radius relation, compactness, and binding energy, are presented in Sec.~\ref{sec:NSP}. In this section, we discuss the essential criteria that the NS configurations must meet to be considered physically realistic, and the constraints imposed by observations (see Sec.~\ref{subsec:MRR}). Furthermore, in Sec.~\ref{subsec:CBE}, we propose a fit for the binding energy of the star with anisotropy. Finally, we give our concluding remarks in Sec.~\ref{discussion}.

\section{\label{sec:NEM} Neutron Star Matter}

\subsection{\label{subsec:TOV}Tolman-Oppenheimer-Volkoff Equations} 

Let us consider an anisotropic fluid in a spherically symmetric spacetime, whose line element is given in terms of the components of the metric $g_{\alpha \beta}$ tensor by
\begin{eqnarray}
\mathrm{d}s^2 &=& -c^{2}\alpha^{2} \mathrm{d}\mathrm{t}^{2} + \left(1 - \dfrac{2Gm}{c^{2}r}\right)^{-1} \mathrm{d}r^2 + r^2 \mathrm{d}\Omega, \label{ds2}
\end{eqnarray}
being $\alpha = \alpha(r)$, $m = m(r)$, $\mathrm{d}\Omega=\mathrm{d}\theta^2 + \sin^2\theta \ \mathrm{d}\phi^2 $, $G$ the gravitational constant and $c$ the speed of light.  The general energy-momentum tensor for a static and spherically
symmetric fluid can be written as~\cite{misner1973}
\begin{eqnarray}
T_{\alpha \beta} &=& (\epsilon + P_\perp) u_\alpha u_\beta + P_\perp g_{\alpha \beta} + (P - P_\perp) n_\alpha n_\beta, \label{T_desc}
\end{eqnarray}
where $P$ is the radial pressure and $P_\perp$ is the tangential pressure. It is worth mentioning that $u^{\alpha}$ and $n^{\alpha}$, given by the following expressions  
\begin{eqnarray}
u^\alpha &=& \left[\dfrac{1}{c\alpha}, 0, 0, 0\right], \\
n^\alpha &=& \left[0 , \left(1 - \dfrac{2Gm}{c^{2}r}\right)^{1/2}, 0, 0\right],
\end{eqnarray}
correspond to unitary time-like and space-like vectors, that is, $u^\alpha u_\alpha = -1$ and $n^\alpha n_\alpha = 1$. 

By solving the Einstein field equations and matter equations, a general expression for an anisotropic spherically symmetric compact star is obtained
\begin{eqnarray} 
\dfrac{\mathrm{d}m}{\mathrm{d}r} &=& 4\pi r^{2}\epsilon , \label{eq:TOV_a}\\
\dfrac{\mathrm{d}P}{\mathrm{d}r} &=& -\dfrac{\left(\epsilon + \dfrac{P}{c^{2}}\right)\left( m + \dfrac{4\pi r^3 P}{c^2}\right)}{\dfrac{r^{2}}{G}\left(1-\dfrac{2Gm}{rc^2}\right)} + \frac{2}{r}(P_\perp - P), \label{hyd}\\
\dfrac{1}{\alpha}\dfrac{\mathrm{d}\alpha}{\mathrm{d}r} &=& \dfrac{G}{c^{2}r^{2}} \left( m + \dfrac{4\pi r^3 P}{c^2}\right)\left(1-\dfrac{2Gm}{rc^2}\right)^{-1} \label{eq:TOV_b}.
\end{eqnarray}
Notice that Eq.(\ref{hyd}) is the only one that contains the contribution of the radial and tangential pressure by the difference $P - P_{\perp}$.  The assumption of spherical symmetry holds for static matter sources where the energy-momentum tensor exhibits the specific condition: $\lvert T_{\theta}^{\theta} - T_{\phi}^{\phi}\rvert << T_{\theta}^{\theta} $. This assumption works in diverse physical scenarios, including those characterized by a solid core, or the state of pion condensation.

To accommodate the transition between the isotropic and anisotropic regimes, it becomes necessary to introduce a functional form for $P - P_{\perp}$, considering that the specific relationship between energy density and radial and tangential pressures is unknown due to its dependence on microscopic factors. Following the work of  \citet{2011CQGra..28b5009H}, the tangential pressure can be written as 
\begin{equation}\label{eq:Delta_p}
P_{\perp} = P \left( 1 + \lambda_a ~ \frac{2 G m }{c^2 r} \right)\, , 
\end{equation}
where $\lambda_a$ is a parameter controlling the degree of anisotropy. When the anisotropy is due to a condensate phase of pions \cite{1972PhRvL..29..382S}, $0\le \left(P - P_{\perp}\right)/P \le 1$, therefore we could expect that the maximum value for the pressure difference is of the order of unity. Considering that the compactness of NSs ranges from $0.05$ to $0.3$ more or less, the range of values for the anisotropy parameter we will be using for this work is $-2\le \lambda_a \le 2$ \cite{2011CQGra..28b5009H,2012PhRvD..85l4023D}. Among various models in the literature, relation (\ref{eq:Delta_p}) introduces the effects of pressure anisotropy in a phenomenological manner \cite[see e.~g.~][for a covariant relation between $P$ and $P_\perp$]{2019PhRvD..99j4072R}. Additionally, there are also works that introduce  pressure anisotropy  self-consistently as the models  presented in \citep{2022PhRvD.105d4025A} of stars with elastic matter \cite[see also][for a review]{2023arXiv230703146A}.

\subsection{\label{subsec:EOS} Equation of State}

We parameterize the NS EOS using the Generalized Piecewise Polytropic (GPP) fit presented in \citet{Boyle2020}. For the mass density, $\rho$, in the interval between $[\rho_i, \rho_{i+1}]$, the pressure, $P$, and energy density, $\epsilon$, are given by:
\begin{eqnarray}\label{eq:poly_eos}
    P(\rho) &=& K_i\rho^{\Gamma_i} + \Lambda_i\\
    \epsilon(\rho) &=& \frac{K_i}{\Gamma_i+1}\rho^{\Gamma_i} +(1+a_i)\rho - \Lambda_i\, ,
\end{eqnarray}
where the parameters $K_i$, $\Lambda_i$ and $a_i$ are determined by enforcing continuity and differentiability of the energy density and pressure across the dividing densities:
\begin{eqnarray}
    K_{i+1}&=& K_i\frac{\Gamma_i}{\Gamma_{i+1}}\rho^{\Gamma_i-\Gamma_{i+1}} \label{eq:Ki}\\
    \Lambda_{i+1}&=& \Lambda_i + \left(1-\frac{\Gamma_i}{\Gamma_{i+1}}\right)K_i\rho^{\Gamma_i}\label{eq:Lambdai}\\
    a_{i+1}&=& a_i +\Gamma_i\frac{\Gamma_{i+1}-\Gamma_i}{(\Gamma_{i+1}-1)(\Gamma_i)-1}K_i\rho^{\Gamma_i-1}\label{eq:ai}
\end{eqnarray}
This paper explores a broad range of nuclear EOS from different models and  composition. All the EOS are listed in Table~\ref{tab:EoS} in Appendix~\ref{sec:app_A} and were obtained using \textsc{Compose}  \cite{2015PPN....46..633T}. For the low-density crust, we use the fit of the Sly(4) EOS given in \cite{Boyle2020}. More details on the best-fits made with the GPP framework can be found in Appendix~\ref{sec:app_A}. We use these fits to compute anisotropic NS configurations.

\subsection{\label{subsec:NC} Numerical Calculations}

All the simulations are computed using a fourth-order Runge-Kutta integrator in a 1D spherical grid extending from $r = 0M$ to the outer domain boundary, $r_{max} = 100 M$.  To avoid the singular behaviour at $r=0$, we follow the procedure shown in \cite{2012arXiv1212.1421G}, where a Taylor expansion is performed around this point. The resulting approximate regular equations are programmed  at least for the first mesh point located at $r = \Delta r$, being $\Delta r$ the uniform spatial resolution of the grid. On the other hand, the radius $R_{\rm NS}$ of the surface of the star is defined as the radius $r = R_{\rm NS}$, where the mass density $\rho(R_{\rm NS}) = 10^{4}$~ g\, cm$^{-3}$. In this sense, the mass of the configuration is $M=m(R_{\rm NS})$. In particular, this code has been used to constrain several configurations of neutron and quark stars using the GW170817 observation \cite{2020Ap&SS.365...43A}. In addition, we have used this code to  study anisotropic quark stars with an interacting quark equation of state, where the contribution of the fourth-order corrections parameter a4 of the QCD perturbation on the radial and tangential pressure generates significant effects on the mass-radius relation and the stability of the quark star \cite{2019PhRvD.100j3006B}.

\section{\label{sec:NSP} Neutron Stars Properties}

\subsection{\label{subsec:MRR} Mass-Radius Relation}

\begin{figure*}
    \centering
    \includegraphics[width=0.99\textwidth]{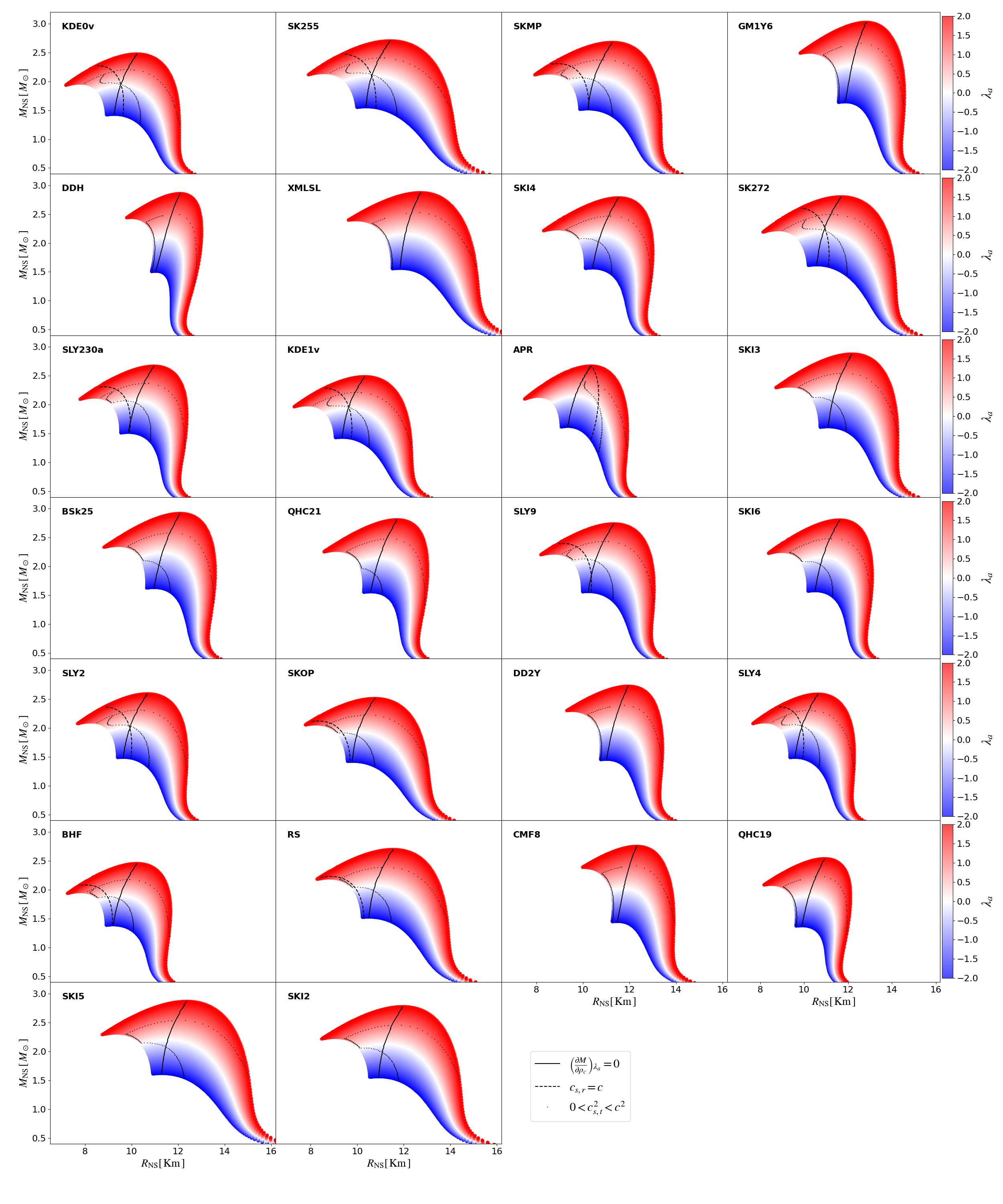}
    \caption{Mass-radius relations for different equations of states with the GPP parameterization (see Table~\ref{tab:EoS}) The color scale corresponds to the value of the anisotropic parameter, $\lambda_a$. Positive values support more massive configurations. Configurations on the left of the solid black line are stable against radial perturbations. Configurations on the right of the dashed black lines satisfy the causality conditions for radial sound speed, while the configurations inside the region enclosed by the dotted black line satisfy it for tangential sound speed.
 }
    \label{fig:MRcs}
\end{figure*}

\begin{figure}
    \centering
    \includegraphics[width=0.95\columnwidth]{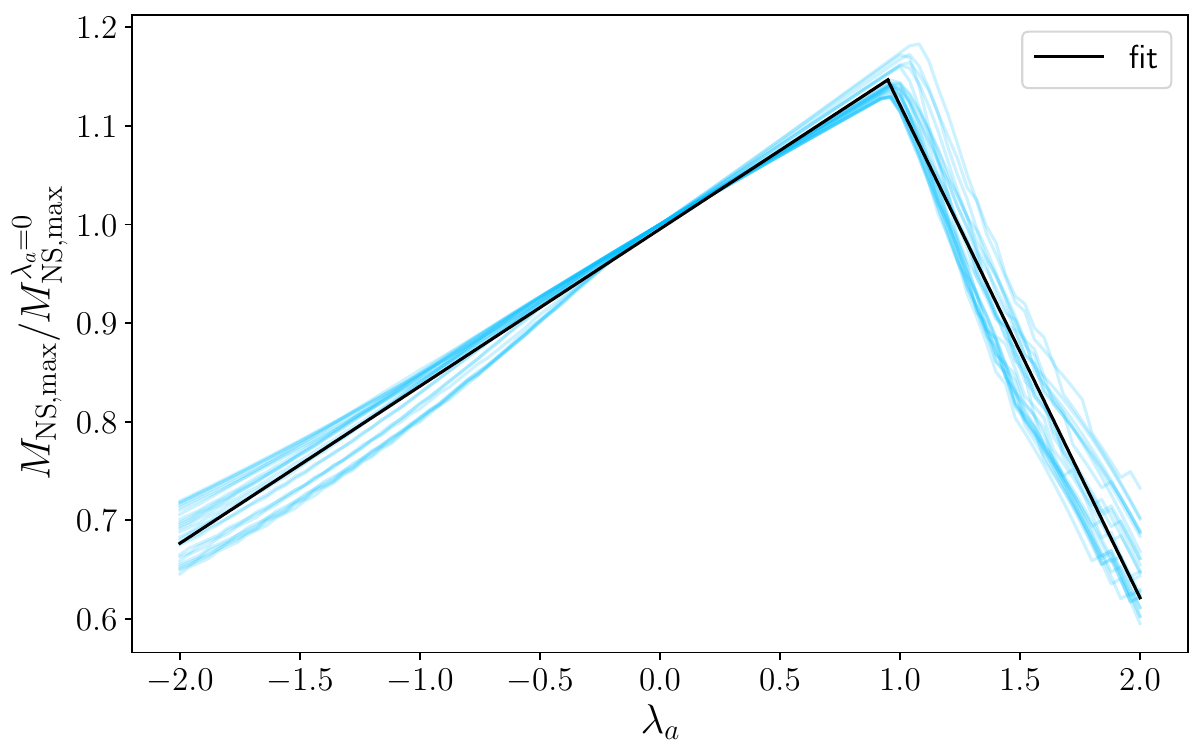}
    \caption{Fraction between the maximum mass for an anisotropic configuration, $M_{\rm NS, max}$, and the isotropic one,  $M^{\lambda_a=0}_{\rm NS, max}$, as a function of the anisotropic parameter. Configuration with $M< M_{\rm NS,max}$ satisfies the stability and causality conditions. Colored lines correspond to the results of the EOS used, and black line is the fit given in Eq.~\ref{eq:fit_mmax}.}
    \label{fig:Mmax_eos}
\end{figure}
For the EOS presented in Table~\ref{tab:EoS}, we solve equations (\ref{eq:TOV_a})-(\ref{eq:TOV_b}) together with equation (\ref{eq:Delta_p}), for different values of the anisotropic parameter $\lambda_a$ and the central mass density $\rho_c$. The resulting mass-radius relations are shown in Fig.~\ref{fig:MRcs}. In general, positive (negative) values of the anisotropic parameter support more (less) massive configurations compared with the isotropic model ($\lambda_a=0$). This is expected since positive (negative) $\lambda_a$ produces an outward (inward) force that is opposite (aligned) to the gravitational force (see Eq.~\ref{hyd}).

We consider the obtained configurations to represent physical objects if they satisfy the two following conditions:
\begin{itemize}
\item Stability condition: the central density $\rho_c$ should be smaller than $\rho_c(M_{\rm max})$, where $M_{\rm max}$ corresponds to the configuration with the maximum mass. Models with $\rho_c > \rho_c(M_{\rm max})$ are unstable to radial perturbation and collapse into black holes \cite{1983bhwd.book.....S}. In Fig.~\ref{fig:MRcs}, configurations located to the left of the solid black line are stable to radial perturbations.

\item Causality condition: The radial and tangential speeds of sound, denoted by:
\begin{equation}
    c_{s,r}^2=\frac{\partial P}{\partial \epsilon} \qquad {\rm and}\qquad c_{s,t}^2=\frac{\partial P_\perp}{\partial \epsilon},
\end{equation}
respectively, must not exceed the speed of light, $c$.  In Fig.~\ref{fig:MRcs}, configurations to the left of the dashed black lines have superluminal radial sound speeds. While, the configurations inside the region bounded by the dotted black line satisfies the condition $0 < c_{\rm s,t}^2 < c^2$. It is important to clarify that in a spherically symmetric anisotropic configuration, up to five independent wave speed modes could be identified \cite{2003CQGra..20.3613K,2023arXiv230703146A}. While  in this work we define and analyze the two main wave modes in the radial and tangential directions, it would be extremely interesting to explore, in future research, whether these configurations exhibit  all wave speeds and if subluminal speeds persist. 
\end{itemize}
Further conditions of acceptability for self-gravitating stellar models have been compiled in other works \cite[see e.g.][]{2022EPJC...82..176S,2023arXiv230706257S}.  But, we only checked for the two above conditions because we considered them to be the minimum set of conditions that a model would have to meet in order to be physically feasible. 

As expected, EOS models built with the relativistic mean field theory, such as  GM1Y6, XMLSL, DDH, DD2Y, QHC19, QHC21,  and  CMF8-EOS, naturally satisfy the casualty condition for the radial speed of sound. This is also the case for the SKI2, SKI3, SKI4, SKI5, SKI6 and BsK-25-EOS. For the BHF, RS, SKMP, SKOP, SLY230a and SLY9-EOS, we found configurations with superluminal radial velocities, but they are unstable to radial perturbations. This is also the case for the remaining EOS studied here, when the anisotropic parameter is positive, $\lambda_a>0$ (i.e.~tangential pressure is greater than the radial one). For the opposite case,  when $\lambda_a<0$ (i.e.~radial pressure is greater than the tangential one), we found stable configurations with radial sound speed greater than the speed of light.  It is particular the case of the APR-EOS because we found this last situation but for all the values of the anisotropic parameter considered here. For this reason, we discard the APR-EOS and do not consider it in the following analysis.

The maximum value of the anisotropic parameter for all EOS is limited by the tangential sound speed. Configurations with superluminal tangential sound speeds occur generally for $\lambda_a>1.0$. In the case of the relativistic EOS, GM1Y6, DDH, XML5L, DD2Y, and CMF8, there are no restrictions on the minimum value of the anisotropic parameter. However, for the remaining EOS, configurations with negative anisotropic parameter can be stable to radial perturbations, but do not satisfy the causality condition of the tangential sound speed, and thus the maximum stable mass is restricted. 

For a given value of the anisotropic parameter, the maximum mass is the mass of the most massive configuration that satisfies both the stability and causality conditions for the radial and tangential sound speeds. Fig.~\ref{fig:Mmax_eos} displays the ratio between the maximum mass for an anisotropic configuration and the one of the isotropic model ($\lambda_a=0$) as a function of the anisotropy parameter, $\lambda_a$, for each EOS.  The maximum mass increases with the anisotropic parameter until $\lambda_a\sim1.0$, where it becomes a decreasing function. In general, for negative values of the anisotropic parameter or values greater than one, the maximum mass is defined by the causality condition for the tangential sound speed,  while for positive values and less than one, it is defined by the stability condition. We can fit the maximum mass with the following relation, independent of the EOS:
\begin{equation}\label{eq:fit_mmax}
M_{\rm NS, max}=  \begin{cases}
M^{\lambda_a=0}_{\rm NS, max}(1.0 + 0.16 \lambda_a)  &\lambda_a< 1.0\\
& \\
M^{\lambda_a=0}_{\rm NS, max}(1.0 - 0.5 \lambda_a)  &\lambda_a> 1.0
\end{cases}
\end{equation}
This relation reproduced the maximum allowed mass for an anisotropic NS configuration with  $1-\chi^2\approx0.015$ for $\lambda_a<1$ and $1-\chi^2\approx 0.038$ for $\lambda_a>1$.
\begin{figure}
    \centering
    \includegraphics[width=0.98\columnwidth]{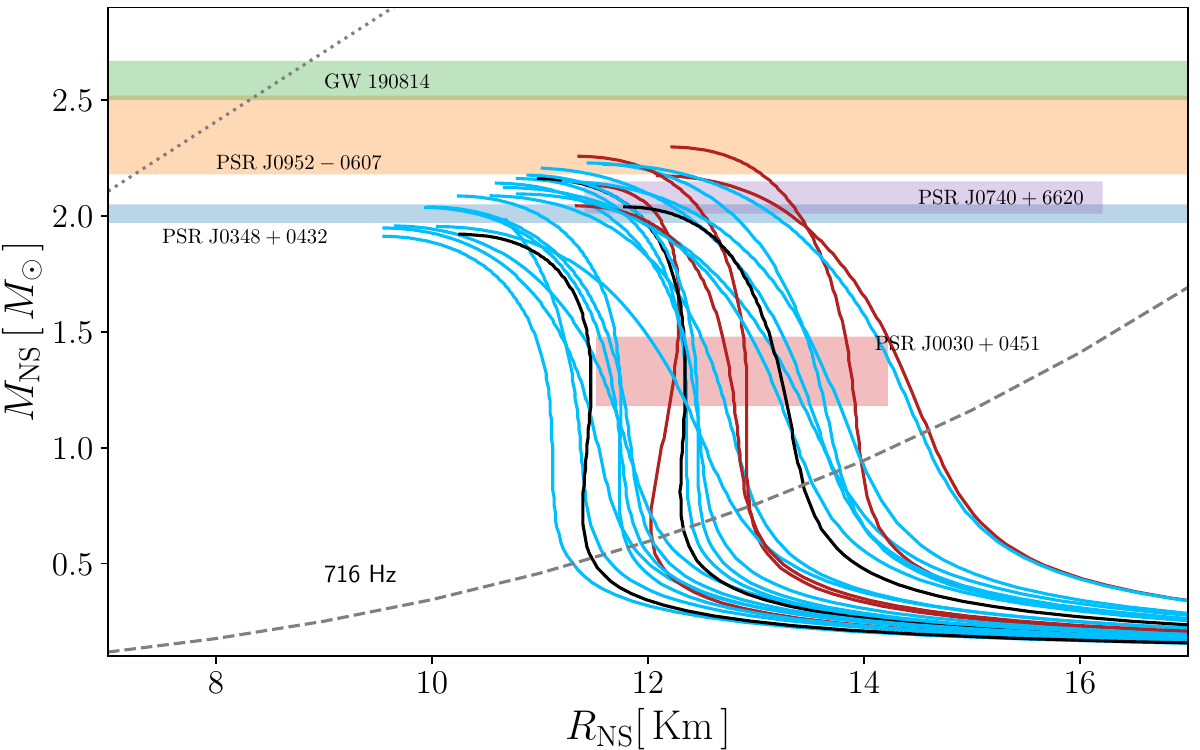}
    \caption{Mass-radius relation of the isotropic configuration for all the EOS summarized in Table~\ref{tab:EoS}. In blue are the configurations composed of npem matter, in red the ones with hyperons and in black the ones with quarks. Colors bands indicate observational constrains given by the pulsar masses of PSR J0348+0432 \cite{2013Sci...340..448A} and PSR J0952-0607, the mass and radius NICER constraints for the pulsars PSR J0030+0451 \cite{2019ApJ...887L..21R,2019ApJ...887L..24M} and PSR J0740+662 \cite{2021ApJ...918L..27R,2021ApJ...918L..28M}, and the mass of the secondary compact object of the GW190814 event \cite{2020ApJ...896L..44A}. The dashed gray line corresponds  to the faster observed pulsar, PSR J1748-2446ad \cite{2006Sci...311.1901H}, and the dotted gray line to the Buchdahl limit for the star compactness \cite{Buchdahl1959}. }
    \label{fig:MR_eos}
\end{figure}
\begin{figure}
    \centering
    \includegraphics[width=0.98\columnwidth]{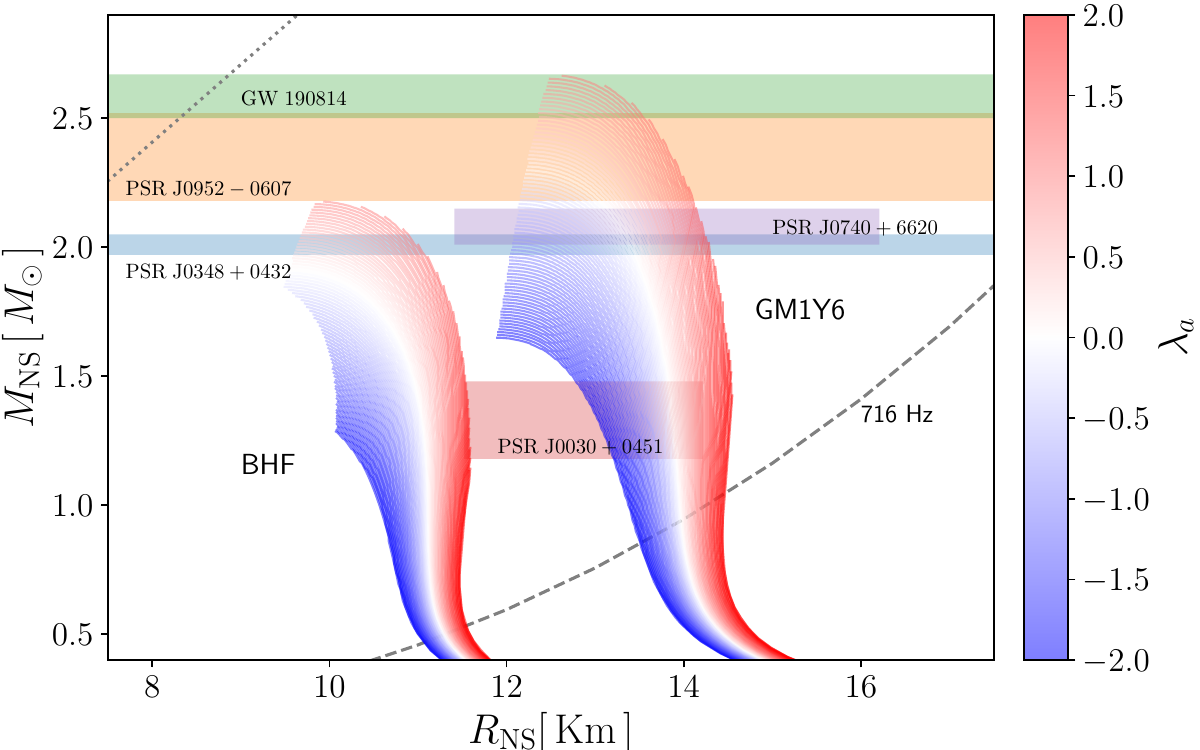}
    \caption{Same as Fig.~\ref{fig:MR_eos} but for anisotropic NSs with, $\lambda_a$ $\epsilon$ $(-2,2)$ for two selected EOS:  BHF and GM1Y6. Only  models satisfying the stability and causality conditions are displayed.}
    \label{fig:MRlambda_eos}
\end{figure}

\begin{table*}
    \begin{tabular}{c|ccccc}\hline
    &  &  &  &  &    \\
    $\quad$ \textbf{EoS} $\quad$ & $\quad$ \textbf{PSR J0348+0432} $\quad$ & \textbf{PSR J0952-0607} $\quad$ & \textbf{PSR J0030+0451} $\quad$ & \textbf{PSR J0740+6620} $\quad$ & \textbf{GW190814} $\quad$ \\ 
    & $M>1.97M_\odot$ & $M>2.17M_\odot$ & NICER  & NICER  & $M>2.5M_\odot$   \\
     &  &  &  &  &    \\ \hline
      &  &  &  &  &    \\
    BHF & $\lambda_a > 0.24$ & $\lambda_a> 0.52$ & \xmark& \xmark & \xmark  \\
    KDE0v & $\lambda_a > 0.08$ & $\lambda_a > 0.88$ & $\lambda_a > 0.24$ &  \xmark  & \xmark  \\
    KDE1v & $\lambda_a > 0.08$ & $\lambda_a > 0.84$ & $\lambda_a > -0.36$ & \xmark &  \xmark  \\
    RS  & $\lambda_a>-0.40$ & $\lambda_a > 0.28$ & \checkmark & $\lambda_a>-0.12$ &\xmark \\
    SK255 & $\lambda_a>-0.40$ & $\lambda_a>0.20$ & \checkmark& $\lambda_a>0.04$ & \xmark \\
    SK272 & $\lambda_a>-0.52$ & $\lambda_a>-0.04$ & \checkmark & $\lambda_a>-0.43$ & \xmark \\
    SkI2 & $\lambda_a>-0.56$ & $\lambda_a>0.12$ & \checkmark& $\lambda_a>-0.40$ & \xmark  \\
    SkI3 & $\lambda_a>-0.76$ & $\lambda_a>-0.12$ & \checkmark& $\lambda_a>-0.63$ & $\lambda_a>0.84$  \\
    SkI4 & $\lambda_a>-0.56$ & $\lambda_a>0.08$ &\checkmark & $\lambda_a>-0.24$ & \xmark  \\
    SkI5 & $\lambda_a>-0.72$ & $\lambda_a>-0.12$  & \checkmark & $\lambda_a>-0.63$ & $\lambda_a>0.88$ \\
    SkI6 & $\lambda_a>-0.60$ & $\lambda_a>0.04$ &\checkmark & $\lambda_a>-0.32$ &\xmark \\
    SkMp & $\lambda_a>-0.36$  & $\lambda_a>0.32$ & \checkmark & $\lambda_a>0.04$ & \xmark \\
    SkOp & $\lambda_a>0.08$ & $\lambda_a>0.8$ & $\lambda_a>-1.68$ & $\lambda_a>0.72$ & \xmark \\
    SLy230a & $\lambda_a>-0.32$ & $\lambda_a>0.32$ & $\lambda_a>-0.76$ & $\lambda_a>0.40$ & \xmark  \\
    SLy2 & $\lambda_a>-0.20$  & $\lambda_a>0.52$ & $\lambda_a>-0.72$  & $\lambda_a>0.76$ & \xmark \\
    SLy4 & $\lambda_a>-0.16$ & $\lambda_a>0.52$ & $\lambda_a>-0.52$ & $\lambda_a>0.80$ & \xmark \\
    SLy9 & $\lambda_a>-0.44$ & $\lambda_a>0.16$  & \checkmark  & $\lambda_a>-0.12$ & \xmark  \\
    BSk25 & $\lambda_a>-0.84$ & $\lambda_a>-0.20$ & \checkmark & $\lambda_a>-0.72$ & $\lambda_a>0.72$ \\
    GM1Y6 & $\lambda_a>-0.92$ & $\lambda_a>-0.32$ &\checkmark  & $\lambda_a>-0.80$ & $\lambda_a>0.56$  \\
    DDH & $\lambda_a>-0.44$ & $\lambda_a>0.16$ & \checkmark  & $\lambda_a>-0.32$ & \xmark  \\
    DD2Y & $\lambda_a>-0.20$  & $\lambda_a>0.43$ & \checkmark  & $\lambda_a>-0.08$ & \xmark
    \\   
    XMLSL & $\lambda_a>-0.60$ & $\lambda_a>0.04$ & \checkmark & $\lambda_a>-0.48$ & $\lambda_a>0.96$ \\
    QHC19 & $\lambda_a>0.16$ & $\lambda_a>0.84$ & $\lambda_a>0.20$ &  $\lambda_a>0.84$ & \xmark \\
    QHC21 & $\lambda_a>-0.56$  & $\lambda_a>0.08$  & \checkmark  & $\lambda_a>-0.32$  & \xmark   \\
    CMF8 & $\lambda_a>-0.20$ & $\lambda_a>0.44$ &\checkmark & $\lambda_a>-0.08$ &\xmark \\
    &  &  &  &  &    \\
    \hline                    
    \end{tabular}
    \caption{ Limiting values for the anisotropic parameter from which the observational constraints are satisfied.  When the ( \xmark ) symbol is present, none of the values within the range $[-2,1]$ satisfy the given constraint. Conversely, when the ( \checkmark ) symbol is present, all configurations within the same range satisfy the constraint. It is worth mentioning that the maximum value of the anisotropic parameter for all EOS is limited by the tangential sound speed. Configurations with superluminal tangential sound speeds occur generally for $\lambda_a > 1.0$.}\label{tab:Pulsars}
\end{table*}

In addition to the conditions mentioned above, we also compare our models to astrophysical observations. Fig.~\ref{fig:MR_eos} shows the mass-radius relation of the isotropic configuration for all the EOS summarized in Table~\ref{tab:EoS}. In this figure,  we have included the following observational constraints:
\begin{itemize}
    \item Observation of pulsar masses:  PSR J0348+0432 with $2.01\pm 0.04~M_\odot$ \cite{2013Sci...340..448A} and PSR J0952-0607 with $2.35\pm 0.17 M_\odot$ \cite{2022ApJ...934L..17R}. 
    \item NICER constraints for the mass-radius relation of PSR J0030+0451 \cite{2019ApJ...887L..21R,2019ApJ...887L..24M} and PSR J0740+662 \cite{2021ApJ...918L..27R,2021ApJ...918L..28M}.
    \item The mass of the secondary compact object of the merger in GW190814 \cite{2020ApJ...896L..44A}, which is between $2.5$ and $2.7~M_\odot$. 
\end{itemize}

As can be seen in Fig.~\ref{fig:MR_eos}, no isotropic model could explain the massive secondary object of GW190814 as a non-rotating NS. On the other hand, BHF, KDE0v, KDE1v-EOS, SKOP and QHC19-EOS would be ruled out by the constraints of pulsar observations.  However, anisotropic configurations have the potential to satisfy some of these constraints for a given EOS, when the isotropic model does not. For example, Fig.~\ref{fig:MRlambda_eos} shows the mass-radius relation with different values of the anisotropic parameter, $\lambda_a$, for the BHF and GM1Y6-EOS.  While the isotropic configurations with the BHF-EOS do not satisfy any of the observational constraints, the anisotropic configurations with $\lambda_a>0.24$ satisfy the mass constraint given by the pulsar PSR J0348+0432, and the configuration with $\lambda_a>0.52$ could model the pulsar PSR J0030+0451. On the other hand, GM1Y6-EOS satisfies nearly all observational constraints, and a configuration with $\lambda_a>0.56$ of the anisotropic model could describe the secondary object of GW190814.

 In Table~\ref{tab:Pulsars} we report the value range of the anisotropic parameter that satisfies the observational constraint for each EOS considered here. If the boundary value of the anisotropic parameter is negative, it means that the corresponding isotropic model satisfies the given constraint. Notably, NICER observations of PSR J0030+0451 rule out BHF, KDE0v, and KDE1v-EOS, even when anisotropic configurations are taken into consideration. The secondary object observed in the GW190814 event could potentially be explained as an anisotropic NS with SKI3, SKI5, GM1Y6, BsK25, and XMLSL-EOS.

\subsection{\label{subsec:CBE} Compactness and Binding-Energy}

For a non-rotating, spherically symmetric NS, its total gravitational mass is defined as:
\begin{equation}
    M_{\rm NS} = \int_0^{R_{\rm NS}} 4\pi r^2\epsilon \, dr \,
\end{equation}
while, its rests mass is given by:
\begin{equation}
    M_{B} = m_B \mathcal{N}
\end{equation}
where $\mathcal{N}$ is the total number of baryons:
\begin{equation}
    \mathcal{N} = \frac{1}{m_B}\int_0^{R_{\rm NS}} 4\pi r^2\rho \left(1-\frac{2Gm}{rc^2}\right)^{-1/2} dr \, ,
\end{equation}
Then, the binding energy of the star, $BE$, is:
\begin{equation}
    {\rm BE}= (M_{\rm NS}-M_B)c^2\, ,
\end{equation}
which can be understood as the amount of energy needed to bring $\mathcal{N}$ baryons together from infinity to form a stable star. During a core collapse, approximately 99\% of the star's binding energy is released through the emission of neutrinos \cite{Lattimer2001}. 

\begin{figure}
    \centering
    \includegraphics[width=0.98\columnwidth]{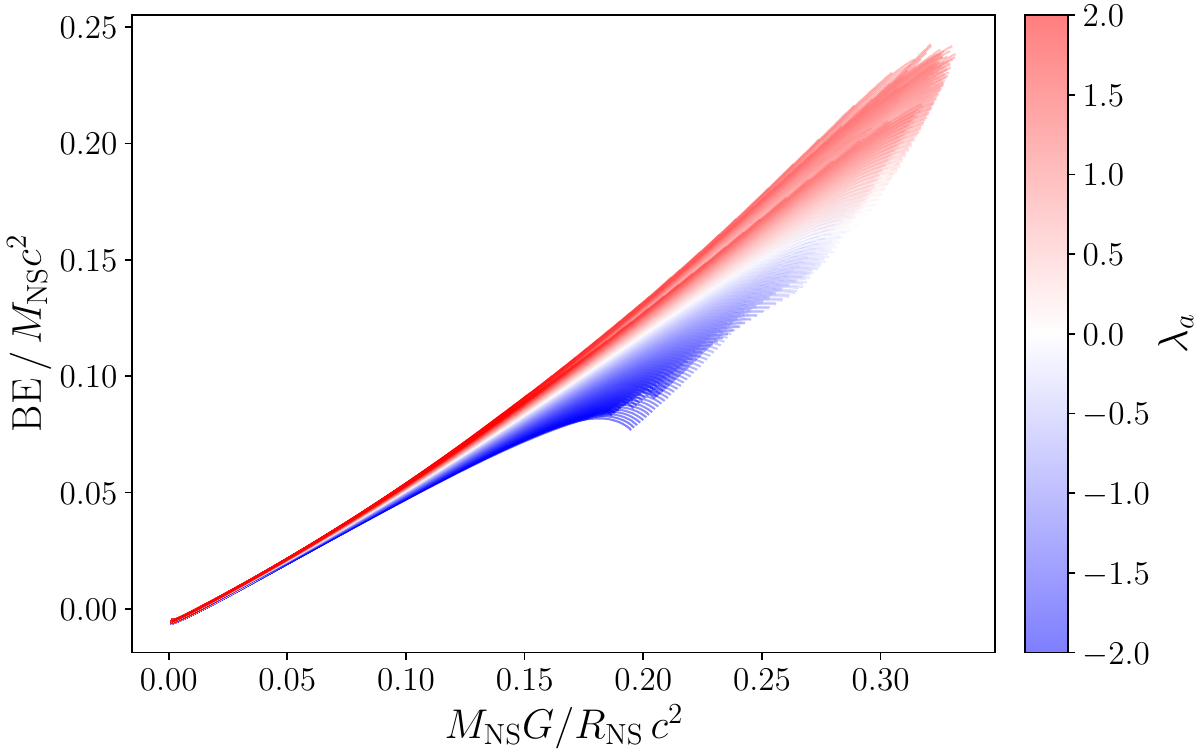}
    \caption{Specific binding energy as a function of star compactness for all EOS summarized in Table~\ref{tab:EoS}. The color scale corresponds to the value of the anisotropic parameter. Only  models satisfying the stability and causality conditions are displayed.}
    \label{fig:BE_eos}
\end{figure}

The binding energy depends on the internal structure of the star and can be calculated using Eqs.~(\ref{eq:TOV_a})-(\ref{eq:TOV_b}).  Fig.~\ref{fig:BE_eos} shows the specific binding energy as function of the compactness of the star, $\mathcal{C}=GM_{\rm NS}/R_{\rm NS}c^2$, for the different EOS and values of the anisotropic parameter. Positive values for the anisotropic parameter increase the binding energy, while, negative values, decrease it. It can be clearly seen that, for the EOS considered here, the specific binding energy is a universal function of the compactness of the star. 

\begin{figure}
    \centering
    \includegraphics[width=0.98\columnwidth]{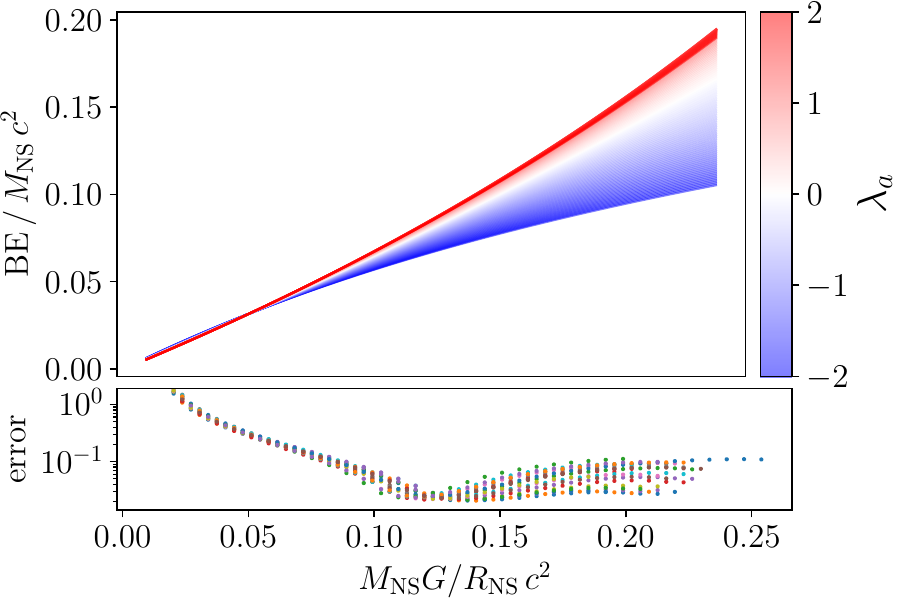}
    \caption{Fit for the specific binding energy as a function of star compactness (see equation \ref{eq:E_binding}). It is worth mentioning that all fits were done up to $\mathcal{C} = M_{max}/R_{max}$.}
    \label{fig:BE_fit}
\end{figure}

In fact, in \cite{Lattimer2001} is proposed a relatively accurate universal relation of the binding energy as:
\begin{equation}
    \dfrac{\rm BE}{M_{\rm NS}c^2} (\mathcal{C})\approx \frac{(0.6\pm0.05)\mathcal{C}}{1-0.5\mathcal{C}}.
\end{equation}
In this work, we propose the following  extended relation that includes anisotropic configurations:
\begin{equation}\label{eq:E_binding}
\dfrac{\rm BE}{M_{\rm NS}c^2} (\mathcal{C},\lambda_a)= \dfrac{\alpha(\lambda_a) \ \mathcal{C}}{1-\beta(\lambda_a) \mathcal{C}} \, , 
\end{equation}
where
\begin{equation*}\label{eq:alpha}
\alpha(\lambda_a)= 0.5213 -0.0197 \lambda_a + 0.0059\lambda_a^2\, , 
\end{equation*}
\begin{equation*}\label{eq:beta}
 \beta(\lambda_a)= 0.5732 + 0.6024 \lambda_a -0.1942\lambda_a^2\, .
 \end{equation*}
The above relation is plotted in Fig.~\ref{fig:BE_fit}. The error reported is calculated as:
\begin{equation}
    {\rm error} = \frac{|{\rm BE}-{\rm BE}_{\rm fit}|}{\rm BE}.
\end{equation}
The differences between the individual EOSs and the universal fits are in the order of a few percent. This universal fit performs better for compactness values greater than $0.1$, with an error less than 10\%.

\section{\label{discussion} Discussions and Conclusions}
In this paper, we study spherical anisotropic configurations of NS, adopting the quasi-local relation for the  tangential  pressure proposed by \cite{2011CQGra..28b5009H} (see Eq.~\ref{eq:Delta_p}). The anisotropy strength is proportional to the anisotropic parameter, $\lambda_a$. It should be noted that our results can be quite sensitive to this choice. We  have studied a wide range of nuclear  EOS and parameterized them using the  GPP formalism \cite{Boyle2020}. We have calculated the mass-radius relation for all of them. As expected, it is affected by the  value of the  anisotropic parameter.  More massive  configurations can be obtained by making the tangential pressure greater than the radial one ( i.e. for $\lambda_a>0$).

We consider an anisotropic configuration to be physically feasible if it is stable to radial perturbations and satisfies the causality condition for the radial and tangential sound speeds. We have found that the maximum value of the anisotropic parameter for all EOS is mainly limited by the tangential sound speed. Configurations with superluminal tangential sound speeds generally occur for $\lambda_a>1.0$ and $\lambda_a<0$.  We have found that introducing anisotropy in the form of Eq.~\ref{eq:Delta_p} can increase the maximum mass for an EOS to about 15\% of the maximum mass of the isotropic configuration.  

In addition, we have compared the mass-radius relations obtained for each EOS with the observational constraints: observations of pulsar masses (PSR J0348+0432 and PSR J0952-0607),  the results of  NICER observations of PSR J0030+045 and PSR J0740+662,  and gravitational wave data. We conclude than an anisotropic configuration has the potential to satisfy the observational constraints, especially if the corresponding isotropic configuration does not (see Table~\ref{tab:Pulsars}). For example, the massive secondary object of GW190814 could be explained by an anisotropic NS.

Finally, we propose a universal relation for the binding energy of anisotropic stars, given by Eq.~\ref{eq:E_binding}. This relation is satisfied for all EOSs used here,  regardless of their particle composition. It is worth saying that this relation, as well as Eq.~\ref{eq:fit_mmax}, depends directly on the anisotropic parameter, $\lambda_a$, introduced in Eq.~\ref{eq:Delta_p}.  For a different function of the tangential pressure, these  fits will not apply.

\begin{acknowledgments}
F.D.L-C was supported by the Vicerrectoría de Investigación y Extensión - Universidad Industrial de Santander, under Grant No. 3703. L.~M.~B is supported by the Vicerrector\'ia de Investigaci\'on y Extensi\'on - Universidad Industrial de Santander Postdoctoral Fellowship Program No. 2023000359. E.~A.~B-V is supported by the Vicerrector\'ia de Investigaci\'on y Extensi\'on - Universidad Industrial de Santander Postdoctoral Fellowship Program No. 2023000354. 
\end{acknowledgments}

\appendix

\section{Parameterization  for the Equation of State }\label{sec:app_A}

In this Appendix, we describe  our numerical approach to obtain a  GPP parameterization for a nuclear EOS.

For the high-density core, following \cite{Boyle2020}, we used a three-segment parameterization, with the limiting densities $\rho_1=10^{14.87}$g~cm$^{-3}$ and $\rho_2=10^{14.99}$g~cm$^{-3}$. For the low-density crust, we use the fit of the Sly(4) EOS given in Table II of  \cite{Boyle2020}. Then, in order to obtain a parameterized model of a given EOS, we need to determine the set of four parameters: $\{\rho_0, \Gamma_1,\Gamma_2,\Gamma_3\}$, where $\rho_0$ is the density separating the star's core from its crust.  The remaining parameters ($K_i, \Lambda_i$ and $a_i$) are given by equations (\ref{eq:Ki})-(\ref{eq:ai}).

\begin{figure}[th!]
    \centering
    \includegraphics[width=0.99\columnwidth]{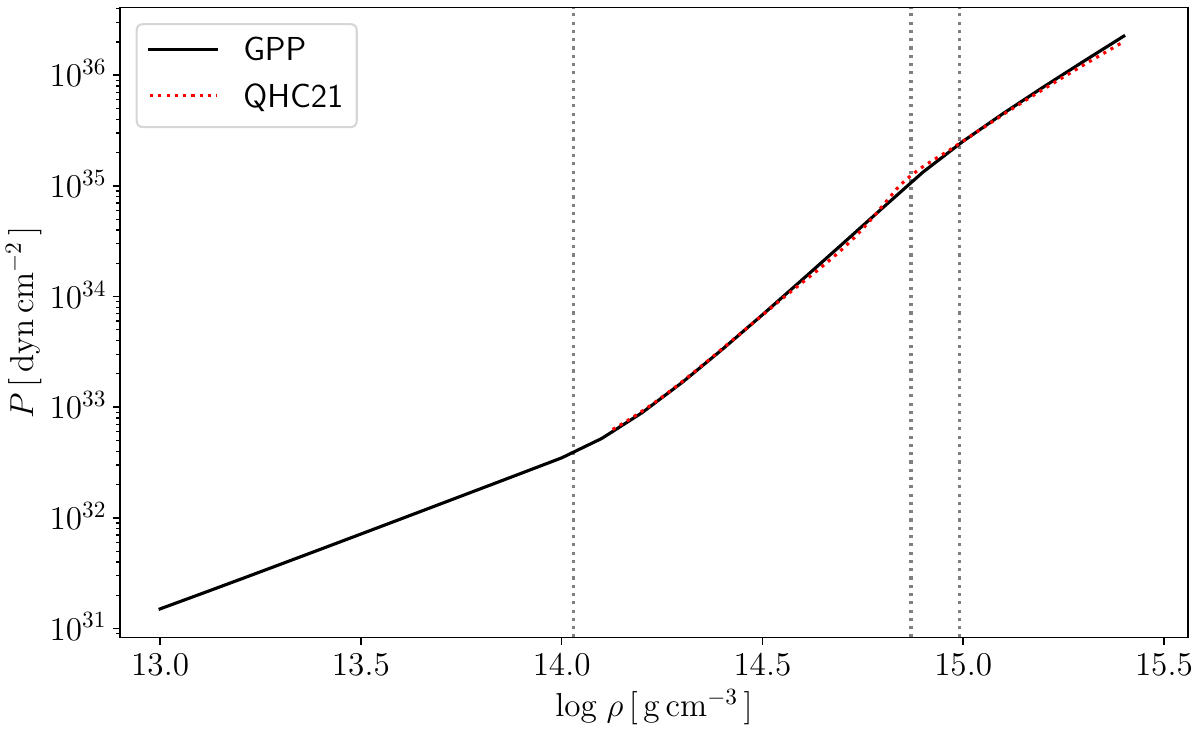}
    \caption{ Pressure as function of mass density for the QHC21 EoS and its corresponding GPP parameterization. The dotted vertical lines correspond to the limiting densities of the  zones shown. }
        \label{fig:fit_eos}
\end{figure}

We perform a Markov Chain Monte Carlo (MCMC) simulation to find the set of parameters: $\{\rho_0, \Gamma_1,\Gamma_2,\Gamma_3\}$ that minimize the difference between the tabulated EOs and the GPP parameterization. For this we used the PyMC3 library \footnote{\url{https://github.com/pymc-devs/pymc}} of Python. The parameters are assumed to be normal distributed. First,  we found an initial guess for the set of  parameters using a Broyden–Fletcher–Goldfarb–Shanno (BFGS) optimization algorithm, and iterate from these values for a total of $40,000$ iterations using a No-U-Turn Sampler (NUTS). The results are summarized in Table~\ref{tab:EoS}. We report the  rms residual of the fit:
\begin{equation}
    \mathrm{rms} = \sqrt{\frac{1}{N}\sum_j (\log P_j - \log(P(\rho_j))^2 }\,
\end{equation}
where $N$ is the number of points of the tabulated EoS, $P_j$ is the tabulated pressure and $P(\rho)$ is given by equation (\ref{eq:poly_eos}). For illustration, Fig.~\ref{fig:fit_eos} shows a comparison of the pressure and mass density relation between the tabulated EOS and the obtained GPP parameterization. 

\begin{table*}
    \begin{tabular}{c|ccccccc}\hline
        &  &  &  &  &  &  & \\
        $\quad$ \textbf{EoS} $\quad$ & $\quad$ \textbf{Composition} $\quad$ & $\mathbf{M_{\rm max}}$ & $\quad$ $ \mathbf{\log \rho_0}$ & $\quad$ $\mathbf{\Gamma_1}$ & $\quad$ $\mathbf{\Gamma_2}$ $\quad$ & $\mathbf{\Gamma_3}$ $\quad$ & $\quad$ $\mathbf{\rm rms}$ $\quad$  \\
          &  & $[\,M_\odot\,]$ & $\quad$ $[\mathrm{g\, cm^{-3}}]$ &  &  &  & \\
         &  &  &  &  &  &  & \\\hline
         &  &  &  &  &  &  &\\
        APR \ \cite{Akmal1998} & npem &  $2.17$ &$14.02\pm 0.02$ & $3.09\pm 0.07$ &$3.59\pm 0.17$ &$3.61\pm 0.20 $ &$0.065$ \\
        BHF \ \cite{Baldo1997}  & npem  & $1.92$ & $14.08\pm 0.02$&$3.21\pm 0.08$ &$2.58\pm 0.18$ &$2.60\pm 0.21$& $0.068$ \\   
        $\quad$ KDE0V \ \cite{Raduta2015,Agrawal2005,Danielewicz2009} $\quad$ & npem &  $1.97$ & $\quad$ $13.95\pm 0.02$ $\quad$ &  $2.92\pm 0.07$ & $\quad$ $2.83\pm 0.19$ $\quad$ & $\quad$ $2.85\pm 0.22$ $\quad$& $0.015$\\
        KDE1V \ \cite{Raduta2015,Agrawal2005,Danielewicz2009}   & npem &  $1.98$& $13.89\pm 0.02$ &$2.84\pm 0.07$ &$2.83\pm 0.19$ &$2.86\pm 0.24$ &$0.022$\\
        RS \ \cite{Raduta2015,Friedrich1986,Danielewicz2009} & npem &   $2.12$ & $13.71\pm 0.03$&$2.73\pm 0.06$ &$2.57\pm 0.28$ & $2.59\pm 0.34$& $0.033$\\
        SK255 \ \cite{Raduta2015,Agrawal2003,Danielewicz2009} & npem &  $2.15$ & $13.65\pm 0.03$ &$2.65\pm 0.05$ &$2.84\pm 0.26$ &$2.85\pm 0.33$& $0.019$\\
        SK272 \ \cite{Raduta2015,Agrawal2003,Danielewicz2009} & npem &  $2.24$& $13.70\pm 0.04$&$2.75\pm 0.06$ &$2.88\pm 0.26$ &$2.93\pm 0.33$ &$0.021$ \\
        SkI2 \ \cite{Raduta2015,Reinhard1995,Danielewicz2009} & npem & $2.17$ & $13.65\pm 0.03$ & $2.70\pm 0.06$ &$2.49\pm 0.26$ &$2.50\pm0.32$& $0.044$\\
        SkI3 \ \cite{Raduta2015,Reinhard1995,Danielewicz2009} & npem & $2.25$ & $13.73\pm 0.03$ &$2.85\pm 0.06$ &$2.43\pm 0.36$ &$2.43\pm 0.42$& $0.046$\\
        SkI4 \ \cite{Raduta2015,Reinhard1995,Danielewicz2009} & npem & $2.18$ & $13.95\pm 0.02$&$3.16\pm 0.07$ & $2.44\pm 0.25$& $2.43\pm 0.29$ & $0.063$\\
        SkI5 \ \cite{Raduta2015,Reinhard1995,Danielewicz2009} & npem & $2.25$ &$13.58\pm 0.04$ &$2.66\pm 0.06$ &$2.54\pm 0.39$ & $2.54\pm0.47$&$0.051$\\
        SkI6 \ \cite{Raduta2015,Reinhard1995,Danielewicz2009} & npem & $2.20$ & $13.94\pm 0.02$ &$3.15\pm 0.07$ &$2.42\pm 0.32$ &$2.41\pm 0.37$& $0.054$\\
        SkMp \ \cite{Raduta2015,Bennour1989,Danielewicz2009} & npem &  $2.17$& $13.80\pm 0.03$ & $2.83\pm 0.06$ &$2.66\pm 0.24$ & $2.67\pm 0.29$ &$0.025$ \\
        SkOp \ \cite{Raduta2015,Reinhard1999,Danielewicz2009} & npem &  $1.98$& $13.77\pm 0.03$ &$2.71\pm 0.06$ &$2.62\pm 0.19$ &$2.64\pm 0.24$ & $0.009$\\   
        SLy230a \ \cite{Raduta2015,Chabanat1998,Danielewicz2009} & npem  & $2.11$ & $14.03\pm 0.02$ &$3.24\pm 0.07$ &$2.63\pm 0.22$ &$2.63\pm 0.26$& $0.032$ \\
        SLy2 \ \cite{Raduta2015,Chabanat1998,Danielewicz2009} & npem &  $2.06$ &$13.96\pm 0.02$&$3.02\pm 0.06$ &$2.79\pm 0.19$ &$2.80\pm 0.23$& $0.005$ \\    
        SLy4 \ \cite{Raduta2015,Chabanat1998,Danielewicz2009} & npem &  $2.06$ &  $13.98\pm 0.02$&$3.05\pm0.07$ &$2.80\pm 0.19$ &$2.81\pm 0.23$& $0.007$\\ 
        SLy9 \ \cite{Raduta2015,Chabanat1998,Danielewicz2009} & npem &  $2.16$ & $13.90\pm 0.02$ & $3.02\pm 0.66$ &$2.66\pm 0.26$ &$2.68\pm 0.31$ & $0.011$\\
        BSk25 \ \cite{Goriely2013,Pearson2018}  & npem  & $2.22$ &$13.90\pm 0.02$&$3.14\pm 0.07$ &$2.26\pm 0.38$ &$2.26\pm 0.43$ & $0.046$ \\
        GM1Y6 \ \cite{Glendenning1991,Oertel2015}  & npeN  & $2.29$ & $13.75\pm 0.02$&$2.96\pm 0.05$ &$1.62\pm 0.35$ &$1.63\pm 0.39$& $0.006$ \\
        DDH \ \cite{Glendenning1991,Gaitanos2004}  & npeN  & $2.05$ & $14.06\pm 0.01$&$3.52\pm 0.05$ &$1.15\pm 0.11$ &$1.14\pm 0.12$&$0.326$ \\
        DD2Y \ \cite{Raduta2020,Typel2010}  & npeN  & $2.04$ & $13.86\pm 0.02$&$2.99\pm 0.06$ &$1.69\pm 0.29$ &$1.69\pm 0.32$ &$0.073$ \\
        XMLSL \ \cite{Xia2022} & npemN & $2.18$ & $13.58\pm 0.03$& $2.68\pm 0.05$ & $1.83\pm 0.47$ & $1.88\pm 0.53$& $0.038$\\
        QHC19 \ \cite{Baym2019}  & npeN-quark &  $1.93$& $14.09\pm 0.02$ &$3.38\pm 0.08$ &$1.95\pm 0.23$ &$1.98\pm 0.27$ &$0.022$ \\
        QHC21 \ \cite{Kojo2022}  & npeN-quark &  $2.20$& $13.98\pm 0.02$ & $3.26\pm 0.07$ & $2.12\pm 0.30$ &$2.22\pm 0.37$ &$0.058$ \\
        CMF8 \ \cite{Dexheimer2008}  & npeN-quark & $2.01$& $13.78\pm 0.03$&$2.89\pm 0.05$ &$1.29\pm 0.22$ &$1.33\pm 0.06$& $0.098$ \\
        &  &  &  &  &  & & \\
        \hline
    \end{tabular}
    \caption{GPP fit parameter for the EOS used in this paper. All EOS used here were obtained with \textsc{Compose} (\url{https://compose.obspm.fr}). npem: neutrons, protons, electrons and muons. npemN: neutrons, protons, electrons and hyperons.} \label{tab:EoS}
\end{table*}

\bibliography{apssamp}

\end{document}